\documentclass[preprint,12pt]{elsarticle}

\usepackage{amssymb}

\journal{Physics Letters B}

\begin{document}

\begin{frontmatter}



\title{Electron screening effects on crustal torsional oscillations}


\author{Hajime Sotani}

\address{Yukawa Institute for Theoretical Physics, Kyoto University, Kyoto 606-8502, Japan.
}

\begin{abstract}
We systematically examine the crustal torsional oscillations as varying the stellar mass and radius, where we take into account the effect of electron screening due to the inhomogeneity of electron distribution. In the examinations, we adopt two different equations of state (EOSs) for the inner crust region of neutron stars. As a result, we find that the frequencies depend obviously on the EOS, even if the neutron star models are almost independent of the EOS. That is, one could solve the degeneracy of the stellar models with different EOS via the observations of shear oscillations frequencies. Additionally, we find that the fundamental frequencies of the $\ell$-th order torsional oscillations can reduce 6\% due to the effect of electron screening, which is independent of the adopted EOS and stellar models. This reduction of frequencies can be crucial to make a constraints on the density dependence of the nuclear symmetry energy, $L$, i.e., the constraint on $L$ can reduce $\sim 15$\% by the electron screening effect.
\end{abstract}


\end{frontmatter}


\section{Introduction}
\label{sec:I}

The structure of neutron stars is still not fixed exactly due to the uncertainty of the equation of stat (EOS) for neutron star matter. This is because the density inside the neutron star can reach up to $\sim10^{15}$ g/cm$^3$ and the examinations of such properties are quite difficult on the Earth.  So, neutron stars may be considered as a suitable laboratory to examine the physics in high density region. In practice, it is suggested that the interior properties of neutron stars can be understood through the observations of oscillations and/or the emitted gravitational waves \cite{AK1996,AK1998,Sotani2001,Sotani2004,GK2011,SYMT2011,SKLS2013}. This is a unique technique known as (gravitational wave) asteroseismology, which is similar to helioseismology for the Sun. Although we still have no direct observations of gravitational waves, we have observational evidences of neutron star oscillations. That is quasi-periodic oscillations (QPOs) in the giant flares from the soft-gamma repeaters (SGRs). SGR is considered as one of the promising candidates of magnetar, which is a neutron star with the magnetic fields stronger than $\sim 10^{14}$ gauss \cite{DT1992}. The extremely strong flare phenomena rarely happen from the SGRs, which are called as the giant flares and different from the usual flare activities. Three giant flares have been detected so far and the QPOs are found in the afterglow of such flare activities. The frequencies of the QPOs in giant flares are in the range from tens of hertz up to a few kilo-hertz \cite{WS2006}.

The discovery of the QPOs in giant flares triggers many theoretical attempts to explain such observations in terms of shear oscillations in the neutron star crust and/or magnetic oscillations (e.g., \cite{Levin2006,Levin2007,Lee2007,SA2007,Sotani2007,Sotani2008a,Sotani2008b,SA2009,Sotani2009}). Through these attempts, it is found that either the crustal torsional oscillations or magnetic oscillations dominate the excited oscillations near the stellar surface, depending on the strength of stellar magnetic fields \cite{CK2011,CK2012,GCFMS2012a,GCFMS2012b}. The magnetic field strength of the SGRs the giant flares arose, may not be so large that the magnetic oscillations become the dominating oscillations, considering the observations of spindown of central objects in the SGRs \cite{K1998,H1999}. Then, assuming that the observed QPO frequencies would come from the crustal torsional oscillations, one can obtain the information about the neutron star matter in the crust region \cite{SW2009,Sotani2011,GNHL2011,SNIO2012,SNIO2013a,SNIO2013b}. Furthermore, one might see the properties about the density region higher than the standard nuclear density through the torsional oscillations \cite{SMT2013}.

To calculate the torsional oscillations, one needs to prepare the shear modulus describing the properties of elasticity, because the restoring force of such oscillations is shear stress due to the elasticity. The principal contribution in the shear modulus must come from the Coulomb energy of the lattice structure composed of nuclei in the crust region. In this context, the shear modulus for bcc lattice in the neutron star crust is derived by Ogata \& Ichimaru \cite{OI1990}. With this formula, many calculations of torsional oscillations have been done as mentioned the above. However, as the secondary contribution in the shear modulus, one should also take into account the inhomogeneous electron distribution in the crust region, i.e., effect of electron screening. In fact, such effects can reduce the shear modulus around 10\% compared to that without such an effect \cite{HH2008,KP2013}. 
In contrast to this simple estimation, the shear modulus strongly depends on the nuclear properties in realistic stellar models, such as the distributions of charge number and Wigner-Seitz radius, and the frequencies of crustal oscillations can be determined by solving the eigenvalue problem in the crust region with the appropriate boundary conditions at the crust basis and stellar surface. This means, it is still uncertain how the frequencies of crustal torsional oscillations depend on the effect of electron screening.

In this article, we examine the frequencies of torsional oscillations with and without the effect of electron screening, as varying the stellar models systematically. Then, we quantitatively evaluate the reduction of frequencies due to such effect,  and discuss the possibility how the previous results should be modified. Probably, this is the first calculations of the frequencies of torsional oscillations in the crust region with the effect of electron screening and we will show the importance of such effect. In particular, we adopt two different EOSs for the inner crust region to see the dependence of EOS \cite{KP2013,DH2001}, where the difference in the adopted EOSs is whether the neutron skin is taken into account or not to produce the crust properties. Identifying the observed QPO frequencies with the crustal torsional oscillations, one may able to get the insight about the neutron skin, which is one of the important properties describing the structure of nucleus in the crust region.
We remark that, according to the macroscopic neutron-star crust models, the non-uniform nuclear structures, the so-called pasta structures, could exist between the crustal region composed of the spherical nuclei and the core region \cite{LRP1993,O1993}, which may play an important role in the crustal torsional oscillations \cite{Sotani2011,GNHL2011}, although the adopted EOSs in this article are not taken into account such properties.

This article is organized as follows. In the next section, we describe the equilibrium configuration of neutron star crust together with the adopted EOSs. In Sec. \ref{sec:III},  we show the equation governing the torsional oscillations and the boundary conditions to determine the eigenfrequencies. Additionally, we also show the obtained spectra of such oscillations. Before concluding, we briefly discuss the importance on the crustal torsional oscillations due to the pasta structures in Sec. \ref{sec:IV}. At the end, we make a conclusion in Sec. \ref{sec:V}. We adopt the geometric unit of $c=G=1$ in this article, where $c$ and $G$ denote the speed of light and the gravitational constant, respectively, and the metric signature is $(-,+,+,+)$.

\section{Crust Equilibrium Models}
\label{sec:II}

We can neglect the magnetic and rotational effects to construct the equilibrium stellar models, because the magnetic energy is much smaller than the gravitational binding energy and the observed magnetars rotate quite slowly. So, we consider spherically symmetric neutron stars in this article, which are given by the solution of the Tolman-Oppenheimer-Volkoff (TOV) equations. The metric of the spherically symmetric spacetime can be written as
\begin{equation}
  ds^2 = -e^{2\Phi} dt^2 + e^{2\Lambda}dr^2 + r^2 d\theta^2 + r^2\sin^2\theta d\phi^2, \label{eq:metric}
\end{equation}
where $\Phi$ and $\Lambda$ are functions of $r$. The function $\Lambda(r)$ is associated with the mass function $m(r)$, such as $e^{-2\Lambda} = 1-2m(r)/r$. In order to close the equation system, one needs to prepare the relation between the pressure $p$ and the energy density $\rho$, i.e., EOS, in addition to the TOV equations. In this article, we adopt the EOS derived by Haensel \& Pichon \cite{HP1994} for the outer crust region. For the inner crust region, we adopt two different EOSs; one is the EOS derived by Kobyakov \& Pethick \cite{KP2013}, which is based on Lattimer \& Swesty's microscopic calculations \cite{LS1991}, and the other is the EOS derived by Douchin \& Haensel \cite{DH2001}. The both EOSs for the inner crust region are derived with the compressible liquid drop model (CLDM) based on the Skyrme-type effective nuclear interaction, but the EOS by Douchin \& Haensel is also taken into account the effect of thickness of neutron skin \cite{DH2001}. Due to the different treatment of the neutron skin, the density at the basis of crust is different from each other. Hereafter, these two EOSs for the inner crust region are referred to as KP2013 and DH2001, and the comparison of two EOSs is shown in Table \ref{tab:EOS}. We remark again that the adopted EOSs, KP2013 and DH2001, do not include the pasta structures at the basis of curst region.

\begin{figure}
\begin{center}
\includegraphics[scale=0.42]{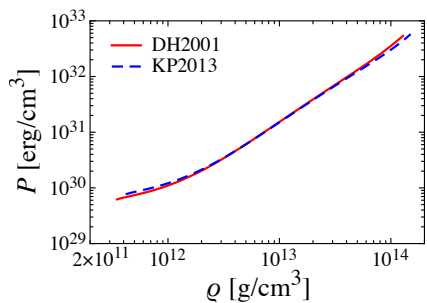} 
\end{center}
\caption{
EOS for the inner crust of neutron star, where the solid and broken liens correspond to the EOS derived by Douchin \& Haensel (2001) and by Kobyakov \& Pethick (2013), respectively.
}
\label{fig:eos}
\end{figure}

\begin{figure}
\begin{center}
\includegraphics[scale=0.42]{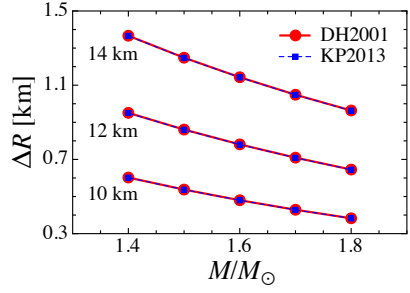} 
\end{center}
\caption{
Crust thickness, $\Delta R$, with different stellar models as a function of the stellar mass, $M/M_\odot$, where the solid lines with circle correspond to the stellar models with DH2001, while the broken lines with square correspond to those with KP2013. The labels denote the corresponding stellar radius.
}
\label{fig:dR}
\end{figure}

\begin{figure*}
\begin{center}
\begin{tabular}{cc}
\includegraphics[scale=0.42]{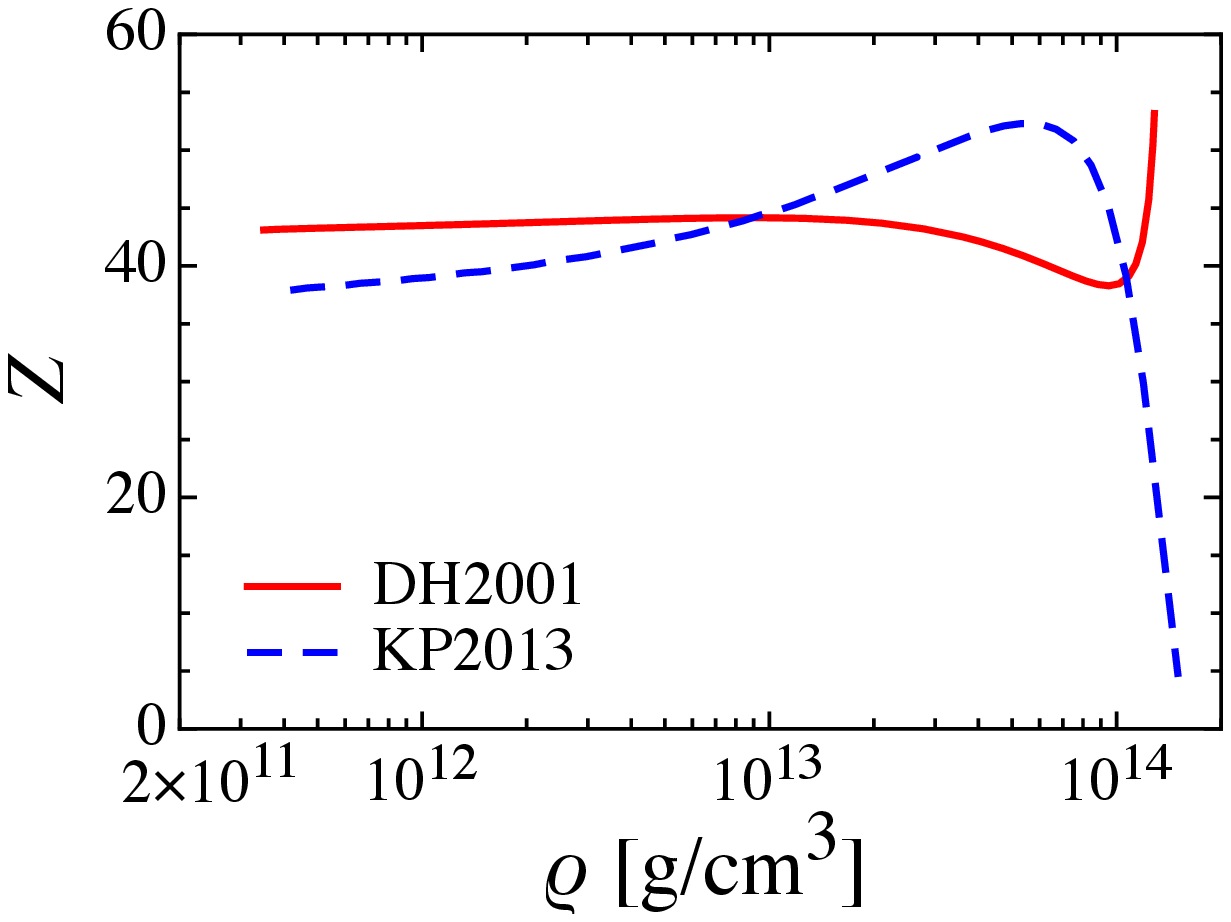} &
\includegraphics[scale=0.42]{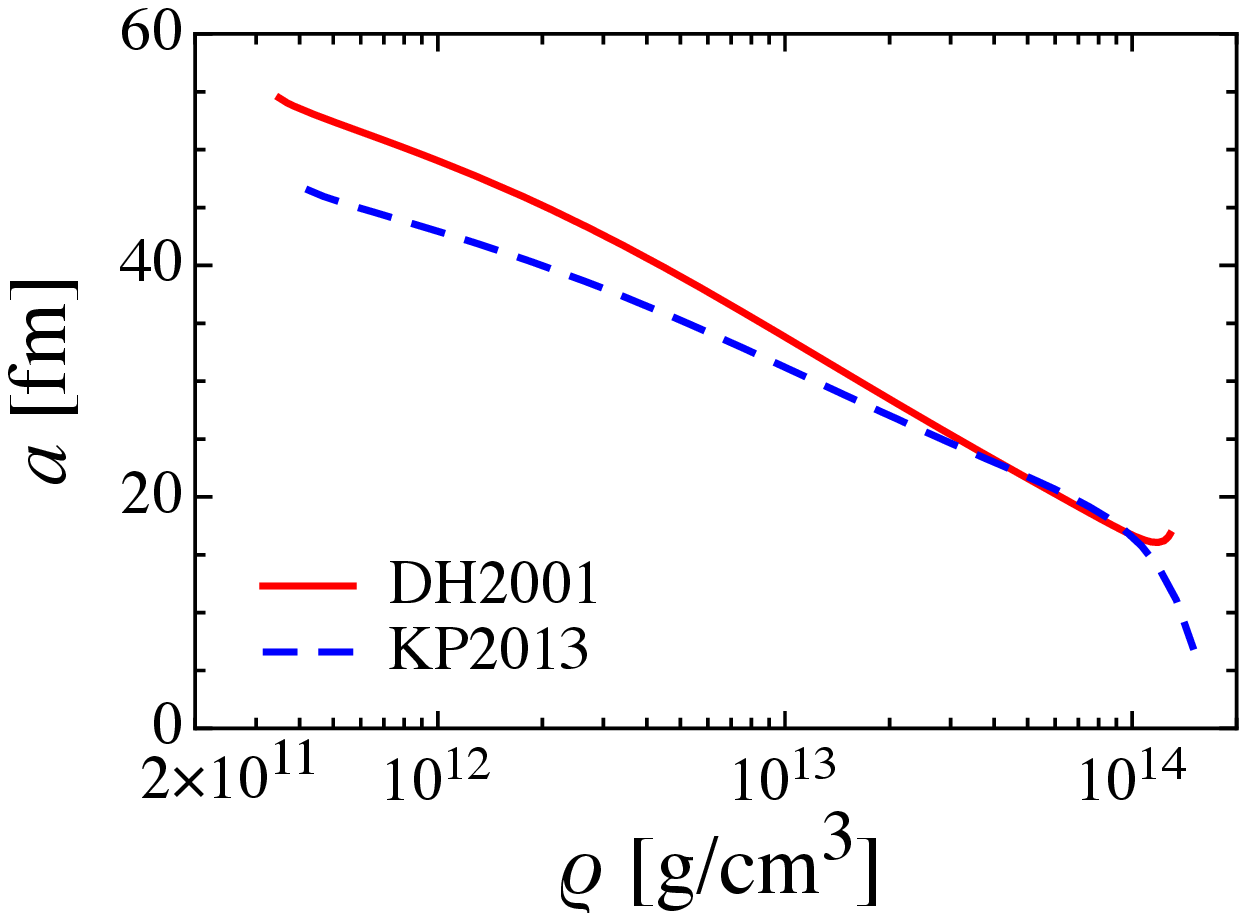} 
\end{tabular}
\end{center}
\caption{
Charge number inside the spherical nuclei, $Z$, in the left panel and the radius of a Wigner-Seitz cell, $a$, in the right panel as a function of the energy density, $\rho$, for the inner crust of neutron star, where the solid and broken liens correspond to the EOS derived by Douchin \& Haensel (2001) and by Kobyakov \& Pethick (2013), respectively.
}
\label{fig:charge}
\end{figure*}

Figure \ref{fig:eos} shows the pressure as a function of the density with two different EOSs for the inner crust region, where one can hardly observe a difference between KP2013 and DH2001. The stellar properties constructed with such EOSs are shown in figure \ref{fig:dR}. From this figure, one can see the degeneracy of stellar models with different EOSs, i.e., it may be impossible to distinguish the crust EOS by only using the direct observations of neutron star itself. However, one can see the difference in the microscopic properties of neutron star matter, such as the charge number, $Z$, (left panel in figure \ref{fig:charge}) and in the radius of a Wigner-Seitz cell, $a$, (right panel in figure \ref{fig:charge}), especially close to the basis of crust. These properties affect on the shear modulus $\mu$, as will be shown later. Consequently, one expects the possibility to distinguish the crust EOS via the observations of the crustal torsional oscillations.

Furthermore, as mentioned before, we focus on how the effect of electron screening affects on the torsional oscillations in the crust region of neutron stars. Thus, we should remove an uncertainty coming from the other factors. In particular, to avoid the uncertainty associated with the EOS for inner core region, we construct the crust equilibrium configuration by integrating the TOV equations from the stellar surface inward up to the basis of crust, as in Refs. \cite{IS1997,SNIO2012,SNIO2013a,SNIO2013b}, with the above EOS for crust region. Then, we will consider the typical neutron star models with $M=1.4-1.8M_\odot$ and $R=10-14$ km.

\begin{table}
\centering
\caption{Comparison between the EOSs derived by Kobyakov \& Pethick (2013) and by Douchin \& Haensel (2001). $n_{bc}$ and $\rho_c$ denote the baryon number density and energy density at the basis of crust.
}
\begin{tabular}{cc|ccc}
\hline\hline
 &  & KP2013 & DH2001 &  \\
\hline
 & model   & CLDM & CLDM &  \\
 & neutron skin   & $\times$ & $\bigcirc$ &  \\ 
 & effective interaction & SI & SLy4  \\
 & $n_{bc}$ [1/fm$^3$]    & 8.913$\times 10^{-2}$ & 7.596$\times 10^{-2}$ &  \\
 & $\rho_c$ [g/cm$^3$]  & 1.504$\times 10^{14}$ & 1.285$\times 10^{14}$ &  \\
\hline\hline
\end{tabular}
\label{tab:EOS}
\end{table}

The shear stress due to the elasticity in the crust region becomes a restoring force for the torsional oscillations, where the shear stress is characterized by the shear modulus $\mu$. The shear modulus in the crust region is mainly determined by the lattice energy due to the Coulomb interaction. In fact, the shear modulus for the bcc lattice is derived by the Monte Carlo calculations averaged over all direction as
\begin{equation}
  \mu = 0.1194\times\frac{n_i(Ze)^2}{a}, \label{eq:off}
\end{equation}
where $n_i$, $Z$, and $a$ are the ion number density, charge number inside the nucleus, and the radius of a Wingner-Seitz cell, respectively \cite{OI1990,SHOII1991}. Most of the previous calculations for torsional oscillations in the crust region have been done with this formula of shear modulus. However, one may have to consider the contribution due to the inhomogeneity of electron distribution, i.e., the effect of electron screening, in the shear modulus. In practice, due to the effect of electron screening, the shear modulus can reduce about 10\% compared to that without such an effect \cite{HH2008}. Recently, the formula of the shear modulus including the effect of electron screening is also suggested as
\begin{equation}
  \mu = 0.1194\left[1-0.010Z^{2/3}\right]\frac{n_i(Ze)^2}{a}, \label{eq:on}
\end{equation}
where the term with $Z^{2/3}$ corresponds to the contribution of the effect of electron screening \cite{KP2013}. With this formula, one can see that the shear modulus reduces $\sim 11.7\%$ for $Z=40$ compared to that without such an effect, which is consistent with the previous suggestion by Horowitz \& Hughto \cite{HH2008}. Furthermore, one might consider the phonon contribution in the shear modulus. But, since such a contribution is much smaller than that coming from a static lattice \cite{B2012}, one can neglect it. Thus, we will calculate the frequencies of torsional oscillations in the crust region with Eqs. (\ref{eq:off}) and (\ref{eq:on}) to examine how important the effect of electron screening is.

\section{Crustal torsional oscillations}
\label{sec:III}

We consider the torsional oscillations on the crust equilibrium configuration mentioned in the previous section. In general, to examine oscillations of neutron stars, one should consider not only the fluid oscillations but also the spacetime oscillations. However, the torsional oscillations are the oscillations with axial parity and do not involve the density variation during the oscillations. Owing to such a feature, one can accurately examine the frequencies of torsional oscillations with the assumption that the metric is fixed during the oscillations, i.e., one can neglect the metric perturbations by setting $\delta g_{\mu\nu}=0$. This treatment is well-known as the relativistic Cowling approximation. Additionally, since the background configuration is spherically symmetric, the non-axisymmetric oscillations degenerate into the axisymmetric oscillations. So, we consider only axisymmetric oscillations in this article. In this case, the non-zero perturbed quantity is the $\phi$-component of perturbed four-velocity, $\delta u^{\phi}$, which can be expressed as
\begin{equation}
   \delta u^{\phi} = {\rm e}^{-\Phi}\partial_t {\cal Y}(t,r)\frac{1}{\sin\theta}\partial_\theta P_{\ell}(\cos\theta).
\end{equation}
In this expression, $\partial_t$ and $\partial_\theta$ denote the partial derivative with respect to $t$ and $\theta$, while $P_\ell$ is the $\ell$-th order Legendre polynomial. Variable ${\cal Y}$ is corresponding to the Lagrangian displacement for the angular direction. Then, the perturbation equation governing the torsional oscillations can be derived from the linearized equation of motion \cite{ST1983} as
\begin{equation}
 {\cal Y}'' + \left[\left(\frac{4}{r} +\Phi'-\Lambda'\right)+\frac{\mu'}{\mu}\right]{\cal Y}' 
      + \left[\frac{\rho+p}{\mu}\omega^2{\rm e}^{-2\Phi}-\frac{(\ell+2)(\ell-1)}{r^2}\right]{\rm e}^{2\Lambda}{\cal Y} = 0,
 \label{eq:perturbation}
\end{equation}
where we assume that the perturbation variable has a harmonic time dependence as ${\cal Y}(t,r)={\rm e}^{{\rm i}\omega t}{\cal Y}(r)$ with eigenfrequency $\omega$.

To determine the frequencies of torsional oscillations, one should impose the appropriate boundary conditions, i.e., the zero-torque condition at the stellar surface ($r=R$) and the zero-traction condition at the basis of crust ($r=R-\Delta R$), because the exterior region of the neutron star is vacuum and the shear modulus in the core region is zero. In practice, since the both conditions can be reduced to ${\cal Y}'=0$ \cite{ST1983,Sotani2007}, we impose such conditions at $r=R$ and $R-\Delta R$. At last, the problem to solve becomes the eigenvalue problem.

\begin{figure*}
\begin{center}
\begin{tabular}{c}
\includegraphics[scale=0.42]{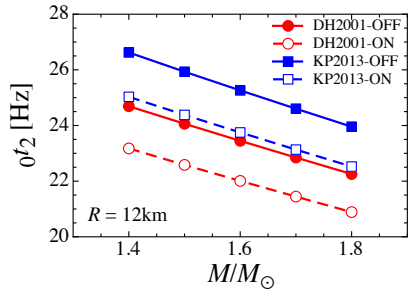} 
\end{tabular}
\end{center}
\caption{
Fundamental frequency of the $\ell=2$ torsional oscillations, ${}_0t_2$, as function of the stellar mass, $M/M_\odot$, for 
$R=12$ km,
where the solid lines correspond to the results without the effect of electron screening, while the broken lines to those with such an effect. The lines with circle are the frequencies with DH2001, while those with square are with KP2013.
}
\label{fig:0t2}
\end{figure*}

\begin{figure}
\begin{center}
\includegraphics[scale=0.42]{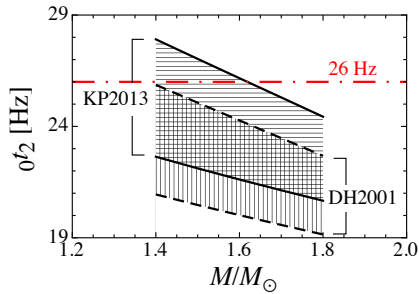} 
\end{center}
\caption{
With the effect of electron screening, the expected fundamental frequencies of the $\ell=2$ torsional oscillations are shown as a function of  stellar mass, where the region between two solid lines corresponds to the expected frequencies for the stellar model with $R=10-14$ km using KP2013, while the region between two broken lines corresponds to those using DH2001. In this figure, one of the QPO frequencies observed in SGR 1806-20, i.e., 26 Hz, is also shown for the comparison.
}
\label{fig:0t2-WB}
\end{figure}

In figure \ref{fig:0t2}, we show the fundamental frequencies of the $\ell=2$ torsional oscillations as a function of the stellar mass with $R=12$ km. In this figure, the solid and broken lines denote the results without and with the effect of electron screening, while the lines with square and circle denote the results with KP2013 and DH2001, respectively. Comparing the solid lines to the broken lines, one can see that the frequencies can reduce 6\% due to the effect of electron screening, which is independent of the adopted crust EOSs and the stellar models. Considering that the lower QPO frequencies observed in SGRs are tens of hertz, this difference due to the effect of electron screening is important to determine the stellar model and/or to obtain the interior information via the QPO frequencies. Additionally, comparing the lines with square to those with circle in this figure, one can observe that the frequencies calculated with DH2001 (lines with circle) become smaller than those with KP2013 (lines with square), where the deviation is around 7\% independent of the stellar models. This is a possible chance to distinguish the crust EOS via the observations of the crustal torsional oscillations, despite a fact that the crust configuration with DH2001 is almost same as that with KP2013 as shown in figure \ref{fig:dR}. Since this difference in frequencies comes from the different treatment of neutron skin for preparing each EOS, one might be possible to get the information about neutron skin via the QPO frequencies from SGRs. In practice, the expected fundamental frequencies of the $\ell=2$ torsional oscillations with the effect of electron screening for the stellar models with $R=10-14$ km can be shown as in figure \ref{fig:0t2-WB}, where the region between two solid lines (shaded with horizontal lines) corresponds to the expected frequencies with KP2013, while the region between two broken lines (shaded with vatical lines)  corresponds to those with DH2001. From this figure, with the help of the other observations of stellar mass and/or radius for central object in SGR, one could be possible to verify the difference of the EOS in the crust region via the QPO frequencies observed in SGRs.

Additionally, we find that the frequencies even for the $\ell=3$ and 4 oscillations reduce 6\% due to the effect of electron screening, which is independent of the adopted crust EOSs and the stellar models as well as the $\ell=2$ oscillations. We also find that the deviation depending on the crust EOS in the frequencies is around 7\% independent of the stellar models even for the $\ell=3$ and 4 oscillations, which is exactly same ratio of the frequency shift as the results for $\ell=2$ oscillations. In other words, one can say that the fundamental frequencies of the $\ell$-th order torsional oscillations can reduce 6\% due to the effect of electron screening independent of the adopted crust EOSs and the stellar models, while the deviation depending on the crust EOS in the frequencies of the $\ell$-th order torsional oscillations is around 7\% independent of the stellar models.


This statement may be powerful to consider the effect of electron screening in the previous results without such effect. For example, the constraints on the density dependence of the nuclear symmetry energy, $L$, with using the QPO frequencies observed in the SGRs suggested in Refs. \cite{SNIO2013a,SNIO2013b} can be shifted to the region with small value of $L$ owing to the reduction of fundamental frequencies. In practice, the allowed value of $L$ from QPO frequencies can reduce $\sim 15\%$, which is quite large modification. This shift may be favorable correction compared to the other experimental results for $L$, which predict a smaller value of $L$  \cite{NGWL2012}.

\begin{figure*}
\begin{center}
\begin{tabular}{c}
\includegraphics[scale=0.42]{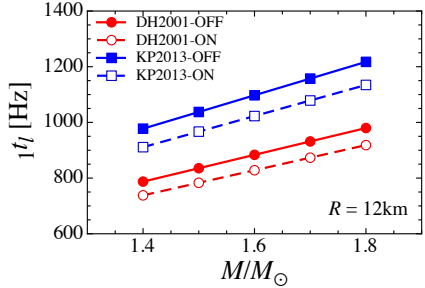} 
\end{tabular}
\end{center}
\caption{
Same as figure \ref{fig:0t2}, but for the frequencies of the 1st overtones of the $\ell$-th order torsional oscillations, ${}_1t_\ell$.
}
\label{fig:1t2}
\end{figure*}

Next, we consider the overtones of the torsional oscillations. First, we confirm that the frequencies of overtones of the torsional oscillations including the effect of electron screening are almost independent of the value of $\ell$, as known in the previous calculations without such effect, i.e., ${}_nt_{2}\cong {}_nt_{3}\cong\cdots\cong {}_nt_{\ell}$, where $n$ denotes the number of nodes in the radial direction. In figure \ref{fig:1t2}, we show the frequencies of the 1st overtone of the $\ell$-th order torsional oscillations as a function of the stellar mass with $R=12$ km. In this figure, as well as figure \ref{fig:0t2}, the solid and broken lines correspond to the results without and with the effect of electron screening, respectively, while the lines with square and circle correspond to the results with KP2013 and with DH2001, respectively. From this figure, one can obviously observe that the frequencies of overtones more strongly depend on the crust EOS rather than the effect of electron screening, unlike the case of fundamental oscillations. In fact, if the stellar model is fixed, the frequencies can reduce 7\% for KP2013 and 6\% for DH2001 due to the effect of electron screening, which is independent of the stellar models. On the other hand, the frequencies deviate around 20\% due to difference of the adopted crust EOS. 
Thus, by identifying the QPO frequencies observed in SGRs with not only fundamental oscillations but also overtones of the torsional oscillations, one could be possible to get an information for stellar properties completely different from that via the other observations, which would help us to understand the physics of neutron stars.

\section{Effect of Pasta Structures}
\label{sec:IV}

Up to now, we consider the torsional oscillations in the crust region composed of the bcc lattice with and without effect of electron screening, where we omit the effect of pasta structures \cite{LRP1993,O1993}. In practice, as increasing the density, the such structures can appear at the basis of crust and the nuclei can form various structures \cite{Okamoto2012,SHHB2013}, which strongly depend on the nuclear symmetry energy \cite{OI2007}. Recently, it is also suggested that the maximum observed spin period of isolated X-ray pulsars could be directly associated with the existence of pasta structures \cite{PVR2013}. Such structures must modify the shear modulus and consequently the frequencies of torsional oscillations. However, the understanding about the shear modulus beyond Eqs. (\ref{eq:off}) and (\ref{eq:on}) is quite poor except for the suggestion that the shear modulus could reduce in the liquid crystal \cite{PP1998}. According to this suggestion, the existence of pasta structures makes the same tendency of the electron screening as shown in this article, i.e., the frequencies of torsional oscillations would reduce due to the pasta structures. Unfortunately, it is difficult to quantitatively estimate the frequency reduction due to the pasta structures, because we have no way how to deal with the shear modulus in the pasta structures. Now, we should emphasize that our finding in this article could be independent of the existence of the pasta structures, and we are groping how to take into account such effects on the shear modulus to examine the oscillations in the more realistic situation.

\section{Conclusion \& Discussion}
\label{sec:V}

The observations of the stellar oscillations of neutron stars are very useful to understand the matter properties in the high density region. In this article, we systematically examine the torsional oscillations in the crust region of neutron stars, where we take into account the effect of electron screening due to the inhomogeneity of electron distribution. In the examinations, we adopt two different EOSs for inner crust region to see the contribution of neutron skin, which is one of the important properties describing the structure of nucleus. As a result, we find that the fundamental frequencies of $\ell$-th order torsional oscillations can reduce 6\% due to the electron screening independent of the adopted EOSs and the stellar models. This uniform shift of fundamental frequencies for all $\ell$ allows us to modify the previous results. In particular, the effect of electron screening can be crucial to constrain the density dependence of the nuclear symmetry energy, $L$, from the QPO frequencies observed in SGRs. For example, the constraint of $L$ in Refs.  \cite{SNIO2013a,SNIO2013b} could be shifted to the region with $\sim 15\%$ smaller values of $L$. 
We remark that we omit the effect of pasta structures because such effect on the shear modulus is still unclear, which should be taken into account somewhere. 



The SGRs accompanying the QPO frequencies are only a few up to now, but one will be able to understand more details of neutron star matter after collecting the observational evidences and identifying those with the torsional oscillations with different $\ell$ together with the additional observations for stellar mass and/or radius. In particular, we emphasize that one can see the difference in the frequencies of the torsional oscillations, even if the neutron star models are almost independent of the EOS. 
That is, the information obtained from the analysis of stellar oscillations will tell us the ``invisible" properties of neutron star matter.

We are grateful to D. Kobyakov for preparing the EOS table and also to K. Iida and T. Tatsumi for their fruitful comments. 
This work was supported in part by Grants-in-Aid for Scientific Research on Innovative 
Areas through No.\ 24105001 and No.\ 24105008 from MEXT, by Grant-in-Aid for Young Scientists (B) through No.\ 24740177 from JSPS, by the Yukawa International Program for Quark-hadron Sciences, and by the Grant-in-Aid for the global COE program ``The Next Generation of Physics, Spun from Universality and Emergence" from MEXT.









\end{document}